# The Microscopic Structure of Stacking Faults in $Sr_2NaNb_5O_{15}$

Robin Sjökvist[1], Yining Xie[1], Zabeada Aslam[2,3], Andy P. Brown[2,3], Nicholas C. Bristowe[4], Mark S. Senn[5], Richard Beanland[1]

[1]Department of Physics, University of Warwick, Coventry CV4 7AL, United Kingdom
[2]School of Chemical and Process Engineering, University of Leeds, Leeds LS2 9JT, United Kingdom
[3]Bragg Centre for Materials Research, University of Leeds, Leeds LS2 9JT, United Kingdom
[4]Centre for Materials Physics, Durham University, Durham DH1 3LE, United Kingdom
[5]Department of Chemistry, University of Warwick, Coventry, CV4 7AL, United Kingdom

## Abstract

Stacking faults and other topological defects in ferroics can have a significant influence on the electronic and mechanical properties of the material. Here, regular stacking faults in the tetragonal tungsten bronze material $Sr_2NaNb_5O_{15}$ are investigated through transmission electron microscopy, symmetry mode analysis and machine-learned force-field calculations. It is shown that the faults, with a fault vector of $\frac{1}{4}[\bar{2}12]_o$, annihilate in sets of four in the material, owing to the ¼ unit cell displacement along the b-axis. The four resulting domains emerge as four possible directions of the $S_3$ order parameter, related to $NbO_6$ octahedral tilts in the material. Force-field calculations reveal that the stacking faults are likely placed at positions where the octahedra in neighbouring domains have similar magnitudes of rotation, and that the estimated stacking fault energy is 46 mJ/m$^2$. The investigation shows that the stacking faults have a significant local effect on the polar modes present in the structure, and therefore could affect the ferroelectric properties.

## Introduction

The tetragonal tungsten bronze (TTB) $Sr_2NaNb_5O_{15}$ (SNN) is currently being investigated as a modern ferroelectric material with application as a dielectric in high temperature capacitors [1–4]. Its high and stable dielectric response across a large temperature range makes it an attractive alternative to polymer-based dielectrics, and is a necessity for efficient energy storage and retention at elevated temperatures [5]. Another benefit is that SNN is Pb and Bi free, making it more compatible with current Ni-based multilayer capacitor electrode technology, unlike many other ceramic oxide ferroelectrics [6,7].

The TTBs have the general formula $(A1)_2(A2)_4C_4(B1)_2(B2)_8O_{30}$, and constitute the second largest group of ferroelectric materials after the perovskites [7]. The large unit cell, in combination with the many different sites suitable for different ionic valencies, makes the basic TTB structure incredibly versatile. In the case of SNN, the A-sites are occupied by Sr and Na, while the B-sites are occupied by Nb and the C-site is vacant. The crystal structure was recently solved through the means of a combination of Transmission Electron Microscopy (TEM), powder X-ray Diffraction (XRD) and neutron diffraction, where it was established that SNN at room temperature adopts an incommensurate, octahedrally tilted structure which approximates closely to an orthorhombic *Ama2* phase [8]. The structure exhibits an out-of-plane polarisation, owing to the displacement of Nb atoms within $NbO_6$ octahedra, which results in the ferroelectric properties. *Ama2* relates to the high temperature tetragonal aristotype *P4/mbm* phase via the basis transformation {(1,−1,0),(2,2,0), (0,0,2)}, and the same crystal structure has also been suggested for the related TTBs $Ba_{0.61}Pb_{0.39}Nb_2O_6$ and $Sr_{0.61}Ba_{0.39}Nb_2O_6$ [9,10]. In Figure 1 (a), the *Ama2* structure is drawn in the $[001]_o = [001]_t$ projection (*o* = orthorhombic, *t* = tetragonal), while the same structure is shown in the $[100]_o = [1\bar{1}0]_t$ direction in Figure 1 (b).

It is well known that the micro- and nanostructure of materials will ultimately affect the device performance, and the presence of stacking faults and other planar defects in ferroelectrics has been shown to affect, for example, the polarization response [11,12]. Dark field (DF) TEM imaging has shown that SNN contains a high density of planar faults that tend to lie on $(010)_o$ planes, and have provisionally been identified as anti-phase boundaries (APBs) [8]. The faults are expected to arise as a way to accommodate the strain arising from the mismatch between the commensurate and incommensurate lattices during growth. Due to the potential impact on the dielectric properties of these planar defects in SNN, a full characterisation of the structure of the planar faults as well as an investigation of the symmetry relationship between the domains they form is warranted.

In this paper, transmission electron microscopy and electron diffraction has been used to reveal and image these planar crystallographic defects in SNN. By using selected area diffraction on individual grains, domains in the structure can be found and properly described, while the utilisation of high resolution TEM allows a demonstration of the nature of the defects. The investigation shows that the planar defect is a type of stacking fault with a preferred habit plane perpendicular to the *b*-axis of the orthorhombic *Ama2* structure, with fault vector $\frac{1}{4}[\bar{2}12]_o$. DF TEM imaging reveals that the stacking faults always merge in sets of four, similar to that observed for the related TTBs BNN and SKN ($Ba_2NaNb_5O_{15}$ and $Sr_2KNb_5O_{15}$, respectively) [13]. Through the use of ISODISTORT [14,15], it was found that the stacking faults can be understood to arise due to the different symmetry-allowed order parameter directions (OPDs) of the $S_3$ order parameter, which is related to cooperative tilting of $NbO_6$ octahedra, expressed as phase shifts. The faults are then expected to form in regions where the displacive modulation of the NbO$_6$ octahedra tends to zero, as this would minimize the energy penalty. Based on this assumption, a structure encompassing four faults was constructed and simulated using force-fields (machine-learned from Density Functional Theory, DFT), revealing a minimal energy cost of 46 mJ/m$^2$. Finally, we speculate on the implications of the faults, and show through symmetry mode analysis, atomic displacements and DF TEM that the presence of the faults has an effect on the ferroelectric properties of SNN.

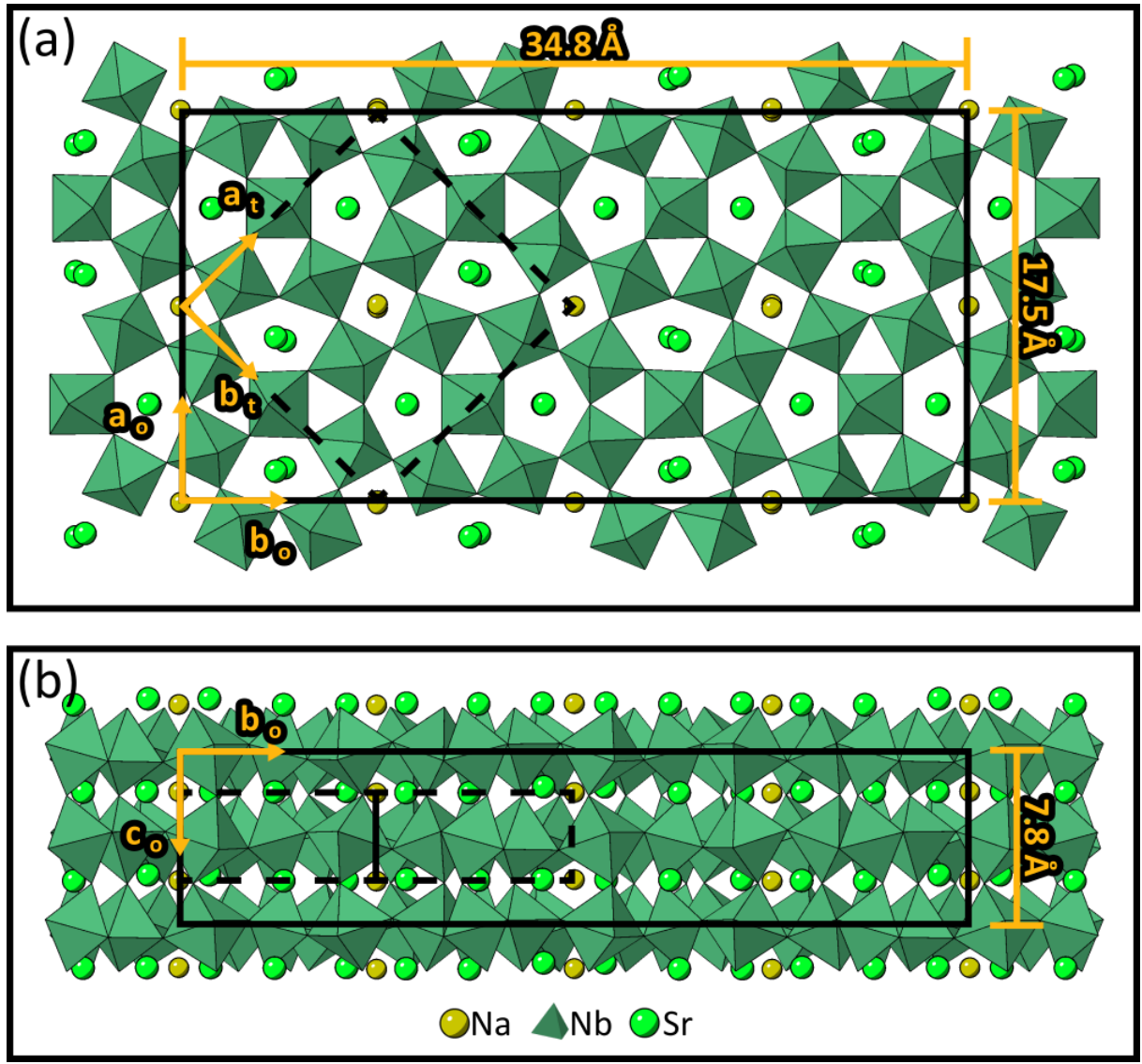

*Figure 1: Crystal structure projections. The orthorhombic Ama2 structure (solid line) viewed along (a) $[001]_o$ and (b) $[100]_o$, with the superimposed extent of the tetragonal aristotype P4/mbm structure (dashed line) viewed along (a) $[001]_t$ and (b) $[1\bar{1}0]_t$. For simplicity we display the structure with 100% occupation of the Sr and Na on the two different A-sites. The O atoms situated at the corners of the Nb-octahedra have been omitted for clarity.*

## Methods

Ceramic powders of SNN were prepared following an established synthesis route [6,8]. To prepare samples for transmission electron microscopy, the ceramic was ground with fine (<10 μm) aluminium powder at a 1:10 ratio and placed in aluminium foil. The folded foil was subsequently pressed using cold rollers, producing a thin solid aluminium sheet containing particles of ceramic which was thinned through mechanical grinding and polishing, mounted to a Cu TEM support using epoxy resin and thinned to transparency using $Ar^+$ ion milling (Gatan PIPS II). High-resolution and dark field/bright field transmission electron microscopy was carried out using a JEOL 2100 $LaB_6$ TEM operated at 200 keV. Atomic resolution annular dark field (ADF) scanning transmission electron microscopy (STEM) was carried out in an aberration corrected JEOL ARM200F FEG TEM operated at 200 keV (20 mrad convergence semi angle and 55-180 mrad detector semi angle). 4D-STEM was caried out in a TESCAN Tensor 100 keV FEG TEM (2 mrad convergence semi angle and 1.66 nm probe size).

In order to tractably, but accurately, simulate the stacking faults in this complex material with a large unit cell, machine-learned force-fields were constructed from DFT (DFT details can be found elsewhere [8]) using the kernel-based "on-the-fly" approach adopted in the VASP software [16,17]. For simplicity, SNN was modelled with 100% occupation of Sr and Na on the two different A-sites (which was previously found to provide good agreement with experiment) [8]. The reference configurations used to build the force-field were generated on-the-fly during a DFT molecular dynamics simulation in which the *Cc* phase (which is similar to the room-temperature *Ama2* structure but with an additional in-plane polarisation, observed to be the ground state of SNN [8]) was slowly warmed up to around 1000K. The accuracy of the force-field was tested against DFT results on structures not explicitly in the training set. Since symmetry breaking from the stacking faults would likely relax the *Ama2* faulted structure to *Cc* (or other low energy phases) in subsequent 0K structural relaxation calculations, the *Cc* phase was used as a model system to simulate the stacking faults.

## Results

Figure 2 (a) illustrates several aspects of the microstructure of SNN. The dark field (DF) transmission electron micrograph highlights two domains of different orientation, with corresponding diffraction patterns in Figure 2 (b) and Figure 2 (c), respectively. The domain corresponding to Figure 2 (b) appears dark and is imaged in the $[010]_o$ zone axis. It shares a curved and uneven border with the second domain, which appears bright and the diffraction pattern of Figure 2 (c) shows that it is aligned to the $[100]_o$ zone axis. The orientational relationship between adjacent domains is therefore that of a 90 degree rotation about $[001]_o$, i.e. orthorhombic twinning. In Figure 2 (c) a spot of the type $\frac{1}{4}(112)_t = (011)_o$, used for the dark field imaging in Figure 2 (a), is indicated by the yellow circle. It is worth mentioning that the spot is displaced slightly from its ideal ¼ index along the orthorhombic b direction, owing to the slightly incommensurate nature of the structure as reported previously [8]. Most notably, these imaging conditions allow the visualisation of bright and dark lines throughout the $[100]_o$ region in Figure 2 (a), corresponding to planar faults in the structure.

Upon closer inspection of the planar faults present in Figure 2 (a), it can be seen that, although they often originate at the border between the twin domains, they tend to run parallel to each other along the c-axis of the orthorhombic cell and only terminate by merging in sets of four. The merging points are expected to be line defects going into the plane (along $[100]_o$) and one such merging point is indicated by a white arrow in the figure. Similarly to the case of hexagonal ErMnO$_3$ where domain walls always meet in sets of six [18], this indicates there is a topological constraint on these crossing points that is determined by the crystallography of the material, as will be discussed later.

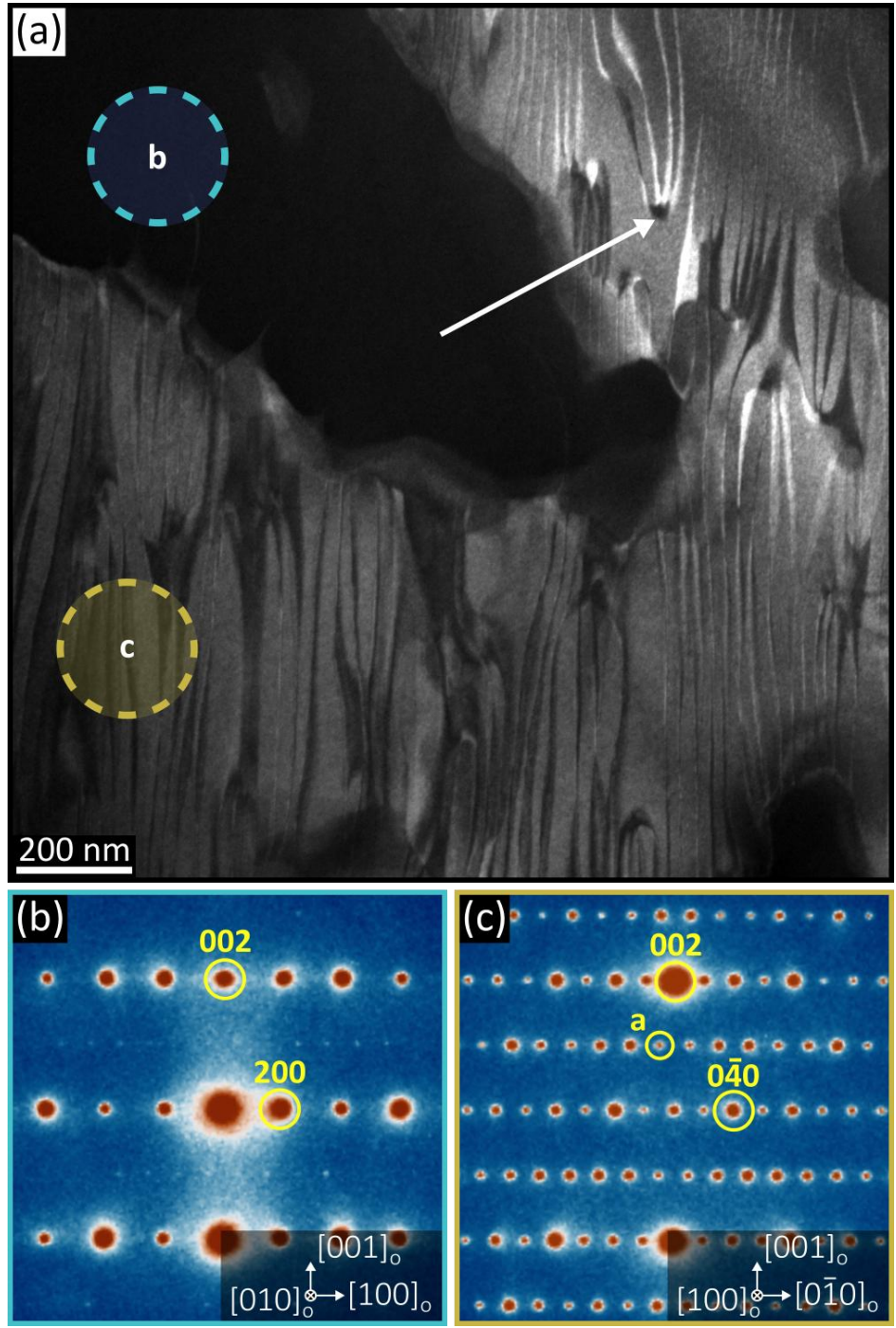

*Figure 2: (a) Dark field micrograph, showing two neighbouring 90° domains, with corresponding diffraction patterns (b) and (c) as labelled in the figure. The spot used when recording the dark field image presented in (a) is indicated by the yellow circle labelled 'a' in (c), and has the indexes $011_o$.*

To obtain a better understanding of the atomistic structure of the faults, the material was also imaged using high-resolution transmission electron microscopy (HRTEM). Unlike aberration-corrected scanning transmission electron microscopy (ac-STEM) these images are dominated by phase contrast that results from the interference of the electron wave passing through the specimen. As such, bright and/or dark spots cannot be associated with individual atom columns and the image does not provide atomic resolution. Nevertheless, the periodicity and symmetry of the image is constrained to match that of the projected crystal potential. Importantly for this material the contrast is exquisitely sensitive to small atomic displacements, including those of light atoms like oxygen, meaning that the periodicity of the *Ama2* structure is readily apparent.

Figure 3 (a) shows a HRTEM image of SNN viewed along the c-axis of the orthorhombic unit cell (same orientation as Figure 1 (a)). Here, the fault is essentially invisible to the uninformed eye but runs vertically through the centre of the image. To demonstrate its presence, two unit cells are outlined in orange on either side of the fault and the resulting fault vector is indicated by a white arrow. A colour scheme is also applied in order to increase the visibility of the features in the unit cells. Looking at the two unit cells drawn to the left of the fault, it can be seen that the contrast of some of the features in the unit cells, highlighted by magenta and yellow arrows, follow a pattern of alternating contrast. Continuing to the dashed "unit cell" drawn in the same row, it can be seen that here the contrast of these features is reversed, indicating that this no longer shows the same unit cell setting. If, however, a shift in the unit cell origin is introduced by the fault vector (described below), the original contrast pattern can be found in the unit cells drawn to the right of the fault. Since the image is sensitive only to projected structure, any shift along the $[001]_o$ viewing direction, i.e. into or out of the image, is invisible. Thus, the fault vector components within the *ab*-plane is seen to be a translation of ½ of a unit cell along *a*, and a ¼ unit cell along *b*.

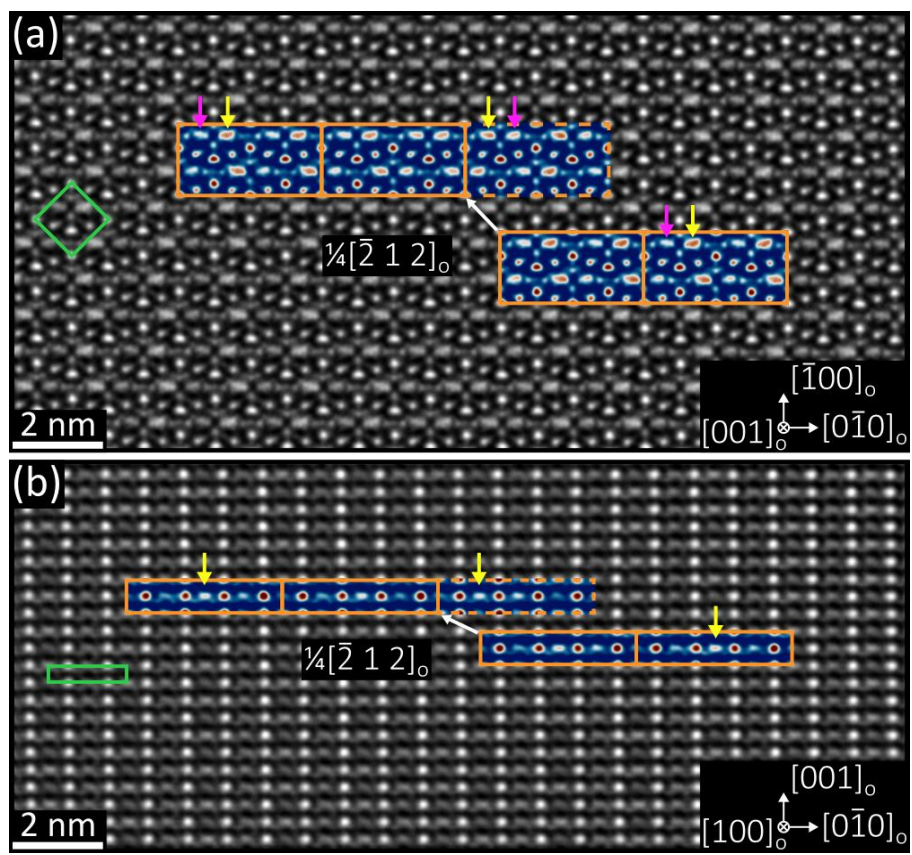

Figure 3: HRTEM micrographs of stacking faults. (a) The structure imaged in the $[001]_o$ direction, giving the in-plane contributions to the fault vector. Yellow and magenta arrows indicate a couple of the features that invert in contrast

*when crossing the fault. Introducing the fault vector restores the contrast. (b) The structure imaged in the $[100]_o$ direction, giving the out of plane contribution to the fault, and a final fault vector of $\frac{1}{4}[\bar{2}12]_o$. The yellow arrow indicates a specific feature of high contrast that shifts when crossing the fault. Its original position is restored by introducing the fault vector. The projection of the aristotype TTB unit cell is indicated in green in both panels. Bragg-filtering is used to improve contrast as shown in the Supplementary Information Figure SI 1.*

The *c*-component to the fault can be found by imaging the material along $[100]_o$, shown in Figure 3 (b) (same orientation as Figure 1 (b)). Here, a planar fault again runs vertically through the image and its presence is demonstrated by the two unit cells marked on either side, with a colour scheme applied. The dashed unit cell clearly shows that the bright feature at the centre of the cell (yellow arrow) ends up in a different spot unless the fault is accounted for. In this projection the $[100]_o$ component of the fault vector cannot be seen, but a shift of ¼ of a unit cell along *b* agrees with that seen in Figure 3 (a), while there is in addition a shift of ½ unit cell along *c*. By combining the results in both Figure 3 (a) and (b), the full fault vector is therefore established as $\frac{1}{4}[\bar{2}12]_o$, as indicated in both panels. Additional atomic resolution STEM imaging was used to confirm the presence of the faults in the $[100]_o$ direction, as shown in Supplementary Information Figure SI 2.

One consequence of the quarter unit cell component in the fault vector is that there are four translational variants that can exist. That is, each defect in the set of parallel stacking faults in Figure 2 (a) shifts the unit cell by $\frac{1}{4}[\bar{2}12]_o$, only returning to the initial translational variant after four stacking faults are passed. This explains the emergence of the four-fault crossing points in the structure. In order to investigate this aspect further, ISODISTORT was used to decompose the distortions related to the breaking of symmetry from the aristotypical SNN structure *P4/mbm* to the present *Ama2* structure [14,15]. It was found that the relevant order parameter giving rise to the domain structure that emerges as a result of the stacking faults transforms as the irreducible representation of $S_3$ (**k** = $\frac{1}{4}[112]$) of *P4/mbm*, which is related to the rotation of the NbO$_6$ octahedra in the structure. This order parameter is formally four-dimensional, arising from a 2-dimensional (little) representation $S_3$ and the two associated arms of the star of **k**. The two arms of the star of **k** correspond to the two possible orthorhombic domains. For a given orthorhombic domain the four high symmetry OPDs associated with just one arm of the star of **k** are illustrated in Figure 4 (a), where the following notation of the permutations is used: $S_3^a$ = $S_3$(a,0;0,0), $S_3^b$ = $S_3$(0,-a;0,0), $S_3^c$ = $S_3$(0,a;0,0) and $S_3^d$ = $S_3$(-a,0;0,0). Alternatively, the different OPDs can be viewed as a simple shift of the unit cell, as is also indicated by the vectors written in Figure 4 (a). Importantly, the symmetry of the space group makes these shifts equal to successive shifts along the fault vector $\frac{1}{4}[\bar{2}12]_o$ that was established from the TEM data and described in Figure 3.

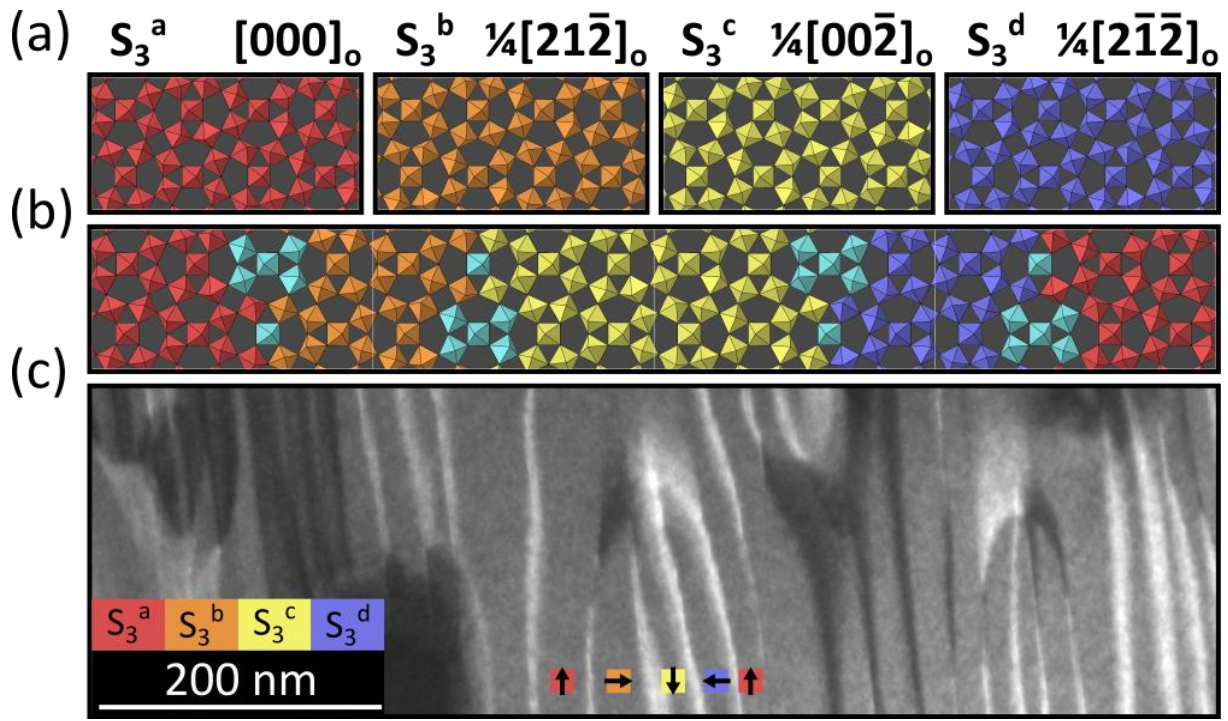

Figure 4: Fault structure. (a) The four different $S_3$ OPDs found, where the differences in tilting patterns of the $NbO_6$ octahedra can be seen. Note that, in order to highlight the octahedral tilting pattern, other order parameters that are needed to fully describe the Ama2 structure are supressed in this representation. (b) A superstructure constructed by interfacing the different $S_3$ OPDs. The cyan octahedra highlight the regions where the tilting pattern is the same for the interfacing OPDs. (c) A DF TEM micrograph, showing several stacking faults. For one set of faults meeting in a four-fold point, coloured dots with added 2D vectors are overlaid to indicate a possible arrangement of the OPDs across the domains.

In order to investigate whether or not it is probable for the four OPDs to form the structure of four stacking faults observed experimentally, and where such faults would occur in the structure, the tilting patterns of the $NbO_6$ octahedra in the OPDs were investigated and compared. It was found that the four OPDs $S_3^a$-$S_3^d$ form a permutation group ($Z_4$), where each OPD contains regions in the unit cell where, locally, the tilting pattern does not change when compared to the tilting pattern of the neighbouring OPDs. For example, the unit cell of OPD $S_3^a$ contains octahedra that have orientations identical to tilted octahedra in $S_3^b$, while a different set of octahedra are equivalent when compared to $S_3^d$. This is illustrated in Figure 4 (b), which is four Ama2 unit cells wide, the smallest supercell that can capture the 4 faults. The four OPDs have been interfaced to one another – colour coded based on the permutations in (a) – with cyan octahedra indicating those octahedra with common tilts. These regions of equivalent octahedra are expected to be low energy locations for stacking faults in the structure. Importantly, the group of OPDs is cyclical, meaning that no overlap in the tilting pattern of $S_3^a$ and $S_3^c$, for example, is found. In Figure 4 (c), the pattern of OPDs is also illustrated by their assigned colours across a four-fold stacking fault structure imaged in DF TEM. Here, 2D vectors have been assigned to the different OPDs based on the cyclical nature of the $Z_4$ permutation group, which forms a pattern reminiscent of a vortex structure or similar topological defect [19].

A deeper investigation of the structural model for the stacking faults was carried out through the means of force-field calculations (machine learned from DFT, see Methods). Although direct comparisons to similar materials are scarce, DFT calculations on stacking faults in perovskites [20] and semiconductors [21], for example, have shown good correlation with experimental investigations. In the present case, the large size of the until cell that describes the *Ama2* structure, even before the stacking faults are considered, makes this work particularly challenging. Here, a structure similar to the one presented in Figure 4 (b) was used as a basis for the force-field relaxation, where the goal was to compare the energy of the structure containing four faults, to a basic structure of similar size without faults. In order to encompass the four stacking faults, a supercell consisting of four unit cells along the orthorhombic b-axis was needed, giving a final length of 139.3 Å. The structure presented in Figure 4 (b) is idealised in the sense that only the $S_3$ order parameter is considered, while the actual structure (and the one used for the force-field calculations) contains additional distortions needed to properly describe the *Ama2* structure of SNN. Furthermore, the structure encompassing the faults was modified to the lower energy *Cc* structure, which the material has been shown to adopt at lower temperatures, since the addition of the faults was expected to break the symmetry of *Ama2* (see Methods) [8]. The resulting structure after relaxation is shown in Figure 5 (a), where the initial placement of the stacking faults is indicated by the dashed lines.

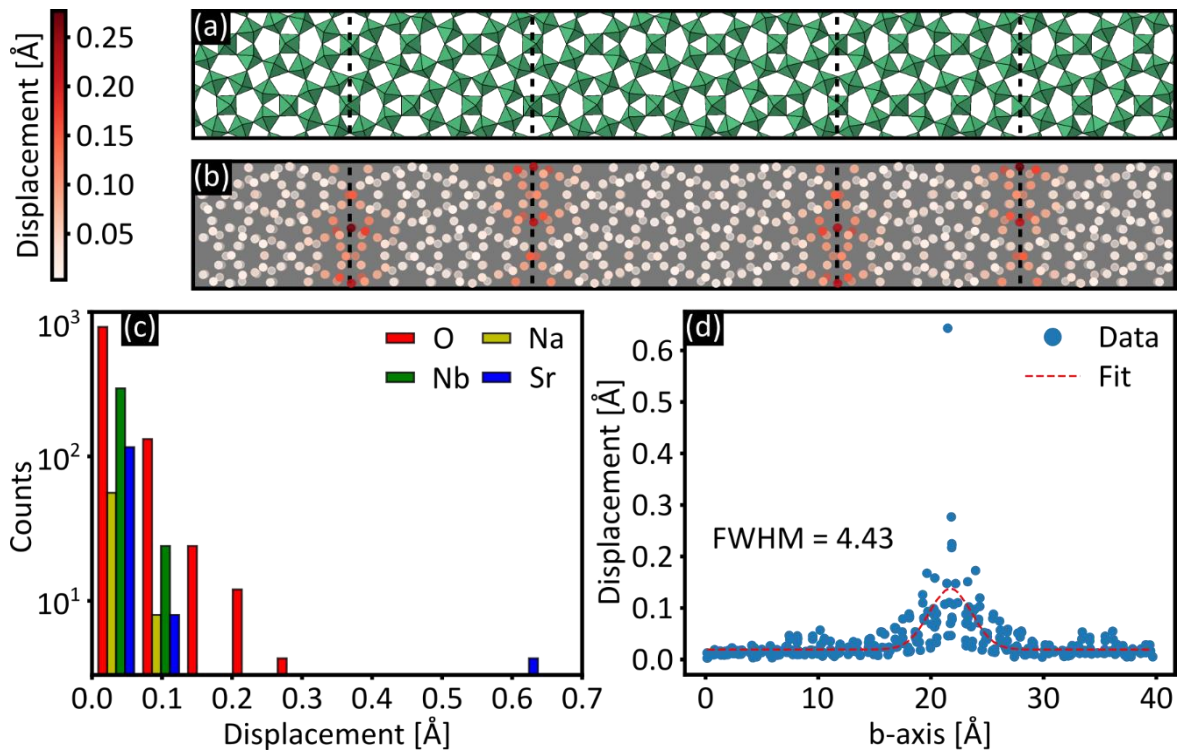

Figure 5: Results from force-field simulations. (a) The structure after relaxation, here only showing NbO$_6$ octahedra. Dashed lines indicate the initial placement of stacking faults. (b) Displacement of O atoms for the relaxed structure with respect to the constructed stacking fault structure, where movement is indicated by the colour scale and initial stacking fault placement is indicated by dashed lines. (c) Histogram (log scale) illustrating the distribution of absolute

*atomic displacements. (d) Scatterplot illustrating all atomic displacements across a stacking fault, along with gaussian fit.*

In Figure 5 (b), the movement of O atoms during the relaxation of the structure is shown, with magnitude indicated by the colour bar. As can be seen, the largest atomic movements are at the stacking faults, suggesting that the initial placement of the faults is a probable low energy position. The movement of all atomic species is illustrated in a similar way on a unified scale in Supplementary Information Figure SI 3. Figure 5 (c) shows the distribution of all atomic movements as a histogram, illustrating that most of the atomic displacements are less than 0.3 Å in magnitude. The exception is a few Sr atoms experiencing displacements of about 0.6 Å. Figure 5 (d) shows a quantification of the displacements around one of the faults, along with a gaussian fit. The fit has a full width at half maximum (FWHM) value of 4.4 Å, giving an estimate of the width of the fault.

The relaxed structure has a 1.57 eV increase in energy compared to a similar *Cc* structure without faults, giving a fault energy of 46 mJ/m$^2$. Such low stacking fault energies are found in, for example, pure metals and alloys [22] (including high entropy alloys [23]) with metallic bonding. In materials with more directional ionic or covalent bonding, low stacking fault energies are only possible when local atomic configurations at the fault are unchanged, for example in rutile $TiO_2$ [24] or tetrahedrally bonded semiconductors including Si [25], and the faults observed here fall into that category. This is not the case for most oxides, including perovskites, which tend to exhibit much larger stacking fault energies [20]. The low stacking fault energy is expected to contribute to the large amount of stacking faults observed in SNN, as shown in Figure 2 (a). The faults occur roughly every 26-29 nm, or about once every 7-8 unit cells along the orthorhombic b-direction (see Supplementary Information Figure SI 4).

The relaxed structure was also compared to the fault-free structure in terms of specific atomic displacements related to the polar nature of the structure. In the *Ama2* setting, the Nb atoms are displaced slightly from the centre of their $NbO_6$ octahedra along the orthorhombic c-axis, giving the material a polar nature, while in the *Cc* setting there is an additional polarisation along the orthorhombic a-axis. The comparison is shown in Supplementary Information Figure SI 5 and 6, and summarised in Table SI 1. The displacement of Nb atoms in the relaxed structure within the width of a singular fault is compared to the displacement in the same area of the fault-free structure, revealing a decrease of 32% along the c-axis and an increase of 50% along the a-axis.

A complementary view was revealed through analysis of the symmetry adapted distortion modes with respect to the aristotype structure. By again cutting up the large supercell with added faults into 20 unit cell sized slabs (with different origin along the b-axis of the supercell), the structures could be decomposed in ISODISTORT and specific mode amplitudes could be compared to those of the *Cc* structure without faults. The variation of the $\Gamma_3^-$ (polarisation along c), $\Gamma_5^-$ (polarisation along a) and $S_3$ (in-plane octahedral tilting) modes is shown in Supplementary Information Figure SI 7 and summarised in Table SI 2. The averaged amplitude of the $\Gamma_3^-$ mode across the unit cells decreased by 21% compared to the fault free *Cc* structure, while the averaged amplitude of the $\Gamma_5^-$ mode increased by 16%. The comparatively small changes revealed by ISODISTORT are related to the size of the system: here, the polar modes are analysed in a larger region than the small area surrounding the faults, giving a lower magnitude overall. Additionally, the amplitude of the $S_3$ mode is suppressed across most of the structure, only reaching values seen for the *Cc* structure in the wide regions between faults. This is expected, since the stacking fault structure was constructed from a combination of different OPDs based on the $S_3$ order parameter and the faults can be seen as interruptions of the octahedral tilting pattern.

TEM images of TTBs can reveal polar domains as well as stacking faults and twin domains [13], as demonstrated in Figure 6 for SNN. The bright field micrograph of Figure 6 (a) shows a grain viewed along $[100]_o$ (c-axis vertical). Contrast due to its internal microstructure is faint, but orthorhombic twin domains and stacking faults, similar to those in Figure 2, are visible. Throughout the image some additional patches of contrast, corresponding to polar domains, can be observed. This additional contrast is due to the influence of $001$ and $00\bar{1}$ reflections, which have different intensities in regions of different c-polarity.

Dark field images – without the strong background of bright field images – have much better contrast, and can be formed using the different reflections visible in the $[100]_o$ diffraction pattern of Figure 6 (c). Figure 6 (d)-(f) show dark field micrographs of one specific region (marked by a dashed square in Figure 6 (b)) using superstructure spots (Figure 6 (d)), a $00\bar{1}$ reflection (Figure 6 (e)) and a $001$ reflection (Figure 6 (f)) respectively. In the latter two images contrast is inverted, indicating a change of polarity along c, i.e. allowing the polar domains in the material to be visualised. They tend to be needle-shaped, with lengths of hundreds of nm along the c-axis while being much narrower along the a-axis (down to the scale of individual unit cells, as shown in the inset of Figure 6 (f)). Notably, there is no correlation between this domain structure and the $\frac{1}{4}[\bar{2}12]_o$ stacking faults in Fig. 6 (d) (this was also investigated and confirmed using 4D-STEM, see Supplementary Information Figure SI 8). In agreement with the force-field simulations, the influence of stacking faults is therefore, at most, a small perturbation of the polar irrep $\Gamma_3^-$. In fact, c-polar domains are even seen to cross twin boundaries (Supplementary Information Figure SI 9). This, together with the needle-shaped domain structure, suggests that the c-axis displacement of Nb in $NbO_6$ octahedra is strongly correlated along $[001]_o$ while being relatively insensitive to structural distortions in other directions [8].

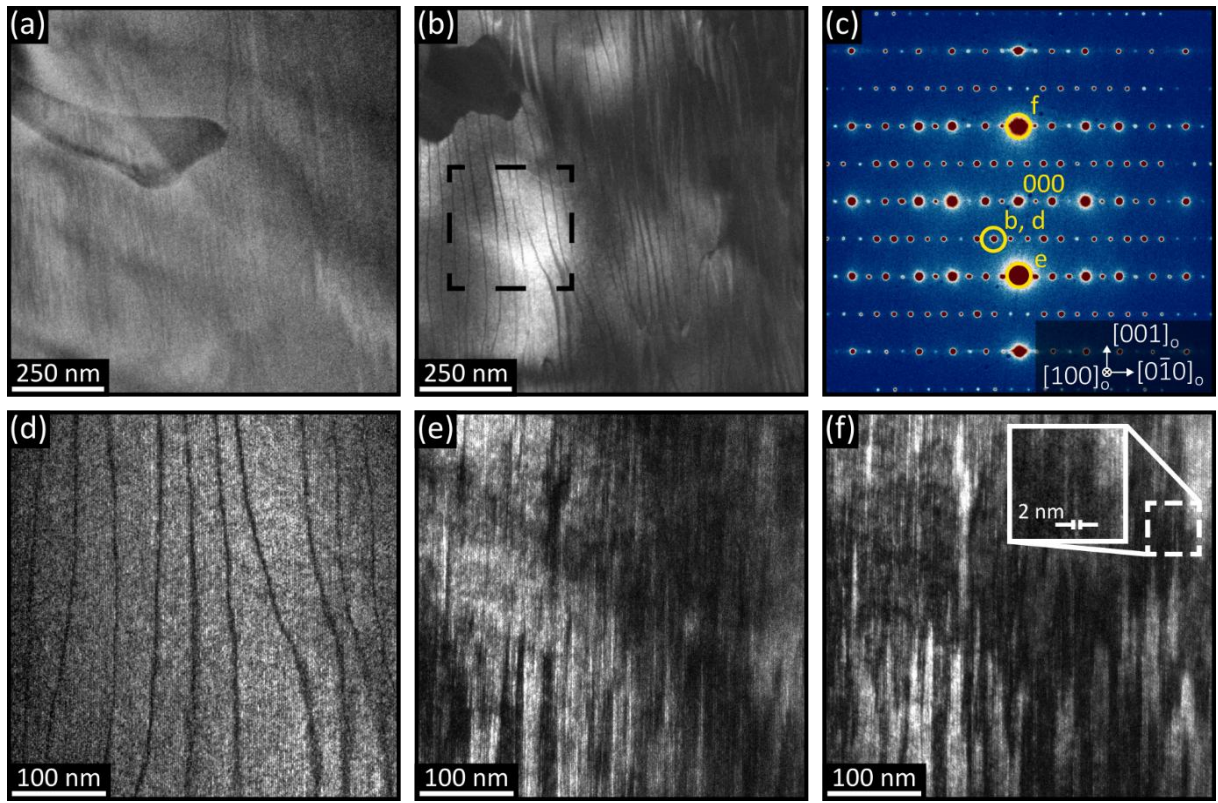

*Figure 6: Investigation of the polarity in SNN. (a) BF and (b) DF of a large area of SNN. (c) Diffraction pattern indicating the spots used for the different imaging conditions. DF patterns highlighting the faults (d), and the extent of opposite polarity (e) and (f), recorded at a higher magnification than (a) and (b).*

## Discussion

Here, observations of microstructure, in combination with machine-learned force-field calculations and symmetry mode analysis, have been used to give insight into the stacking fault structure of SNN. Several factors speak to the validity of the structural model used for the calculations: the small magnitude of atomic movements during relaxation, and the lack of movement from their original location, are both indications of an accurate initial placement of the faults. The estimated stacking fault width of 4.43 Å seems reasonable based on their contrast in HRTEM (Figure 3) and ADF STEM (Supplementary Information Figure SI 2). This small width justifies the relatively small size of the model (only four unit cells) used for the force-field calculations, which is still large enough for the faults not to overlap or interact directly. It can also explain why the faults do not align with the polar domains: the energy penalty for placing a polar domain wall at the same location is expected to decrease for a wider stacking fault, as the domains formed would be less correlated. Furthermore, the low stacking fault energy of 46 mJ/m$^2$ estimated from the calculations agrees well with the small width seen, but also the high density of faults observed in TEM (Supplementary Information Figure SI 4).

Based on the force-field calculations the polarisation of the structure was investigated using two complementary methods, analysing the polar displacement of Nb atoms within $NbO_6$ octahedra and through symmetry-mode analysis in ISODISTORT. Both methods suggest a decrease in the c-axis polar $\Gamma_3^-$ mode at the faults (Supplementary Information Figures SI 5, SI 7 and Tables SI 1, SI 2), where the displacement analysis gives a more localised value. However, 001 DF TEM images show that correlation of the $\Gamma_3^-$ mode along the c-axis dominates above all else, with no observable correlation between the polar $\Gamma_3^-$ domains and stacking faults.

Regarding the $\Gamma_5^-$ mode, giving in-plane polarisation along the orthorhombic a-axis, both Nb displacements and the symmetry mode analysis suggest an increase in the presence of stacking faults (Supplementary Information Figures SI 6, SI 7 and Tables SI 1, SI 2). Given the large increase observed in the displacement analysis, the abundance of faults are expected to have a degree of interplay with the low temperature phase transition from *Ama2* to *Cc* (on cooling), related to the in-plane polar instability [8]. However, further analysis specifically targeting the low temperature phase transition is required in order to prove this connection.

Planar defects similar to those shown here have been described for related TTBs and other functional oxides in earlier studies, where it has also been shown that there is a preference for the faults to meet and annihilate in sets of four [13,26]. The necessity of forming four faults when the displacement vector contains a ¼ in any unit cell direction has also been demonstrated in $Ni_3Mo$ [27]. The meeting points of the faults also show similarities to multiferroic vortex structures such as the 'clover-leaf' pattern found in the hexagonal manganates $YMnO_3$ [28] and $ErMnO_3$ [29], and the TTB $CsNbW_2O_9$ [30]. Currently, domain walls, vortices and other topological structures in ferroelectrics are being investigated for their potential in, for example, memory and logic applications [31,32]. In $YbFeO_3$ the stacking faults act as pinning centres for ferroelectric domains, and therefore, the material on either side of the fault would exhibit inverse polarization [33]. However, as already discussed, there is no such connection between the faults and polar domains in SNN. On the other hand, the fact that the four interfaced $S_3$ order parameter directions form a cyclical $Z_4$ permutation group would indicate that the domains form a non-trivial topological structure reminiscent of a vortex.

It has also been shown previously that the $S_3$ order parameter is correlated with the incommensurate modulation along the b-axis of the orthorhombic cell [8]. Since the structure discussed here is constructed from the order parameter directions of $S_3$, this strengthens the idea that there is a connection between the emergence of stacking faults and the mismatch between the commensurate and incommensurate lattices (see Introduction). A possible route for tuning the ferroelectric properties of these materials is therefore found in controlling the incommensurate modulation, and by extension the stacking fault density.

# Conclusions

In conclusion, stacking faults in Sr$_2$NaNb$_5$O$_{15}$ have been thoroughly described based on a combination of transmission electron microscopy imaging, symmetry investigation through ISODISTORT and machine-learned force-field calculations. The investigation revealed that the stacking faults have a fault vector of $\frac{1}{4}[\bar{2}12]_o$, meet in sets of four, and form domains that can be related as the expression of four directions of the order parameter S$_3$. This information was then used to construct a probable structure for the force-field calculations, which revealed a minimal stacking fault energy of 46 mJ/m$^2$. Finally, the analysis of the force-field calculations indicated that the stacking faults have an effect on the polar modes present in the material, with the strongest effect seen for the in-plane polar mode.

# Acknowledgements

The authors would like to thank T. Brown, C. M. King and S. J. Milne for supplying the SNN ceramic. The authors would like to acknowledge funding from the EPSRC grant "New directions in high temperature dielectrics: unlocking performance of doped tungsten bronze oxides through mechanistic understanding", EP/V053701/1.  M. S. Senn acknowledges support from the Royal Society (URF/R/231012) for a University Research Fellowship. The authors would also like to thank J. P. Tidey for fruitful discussions.

# References

[1]	P. Si, P. Zheng, X. Zhang, C. Luo, X. Zheng, Q. Fan, W. Bai, J. Zhang, L. Zheng, and Y. Zhang, Synergistic modulation of ferroelectric polarization and relaxor behavior of Sr$_2$NaNb$_5$O$_{15}$-based tungsten bronze ceramic, Int J Appl Ceram Technol **21**, 532 (2024).

[2]	Y. Dan, X. Zheng, Y. Meng, S. Wu, C. Hu, L. Liu, and L. Fang, Simultaneously achieving large energy storage density and high efficiency in the optimized Sr2NaNb5O15 system with excellent temperature stability at a low electric field, Ceram Int **50**, 6801 (2024).

[3]	K. Yu, X. Zhang, W. Zhong, P. Zheng, Q. Fan, L. Zheng, Y. Zhang, and W. Bai, Relaxor regulation and enhancement of energy storage properties in bi-modified Sr2NaNb5O15-based tetragonal tungsten bronze ceramics, Journal of Materials Science: Materials in Electronics **34**, 2069 (2023).

[4]	S. Xu, R. Hao, Z. Yan, S. Hou, Z. Peng, D. Wu, P. Liang, X. Chao, L. Wei, and Z. Yang, Enhanced energy storage properties and superior thermal stability in SNN-based

tungsten bronze ceramics through substitution strategy, J Eur Ceram Soc **42**, 2781 (2022).

[5]  S. M. R. Billah, *Dielectric Polymers*, in (2019), pp. 241–288.

[6]  T. Brown, A. P. Brown, D. A. Hall, T. E. Hooper, Y. Li, S. Micklethwaite, Z. Aslam, and S. J. Milne, New high temperature dielectrics: Bi-free tungsten bronze ceramics with stable permittivity over a very wide temperature range, J Eur Ceram Soc **41**, 3416 (2021).

[7]  L. Cao, Y. Yuan, E. Li, and S. Zhang, Relaxor regulation and improvement of energy storage properties of $Sr_2NaNb_5O_{15}$-based tungsten bronze ceramics through B-site substitution, Chemical Engineering Journal **421**, 127846 (2021).

[8]  J. P. Tidey, U. Dey, A. M. Sanchez, W.-T. Chen, B.-H. Chen, Y.-C. Chuang, M. T. Fernandez-Diaz, N. C. Bristowe, R. Beanland, and M. S. Senn, Structural origins of dielectric anomalies in the filled tetragonal tungsten bronze $Sr_2NaNb_5O_{15}$, Commun Mater **5**, 71 (2024).

[9]  V. Krayzman, A. Bosak, H. Y. Playford, B. Ravel, and I. Levin, Incommensurate Modulation and Competing Ferroelectric/Antiferroelectric Modes in Tetragonal Tungsten Bronzes, Chemistry of Materials **34**, 9989 (2022).

[10] T. Woike, V. Petříček, M. Dušek, N. K. Hansen, P. Fertey, C. Lecomte, A. Arakcheeva, G. Chapuis, M. Imlau, and R. Pankrath, The modulated structure of $Ba_{0.39}Sr_{0.61}Nb_2O_6$. I. Harmonic solution, Acta Crystallogr B **59**, 28 (2003).

[11] Y. Ding, J. S. Liu, J. S. Zhu, and Y. N. Wang, Stacking faults and their effects on ferroelectric properties in strontium bismuth tantalate, J Appl Phys **91**, 2255 (2002).

[12] H. Ding, N. Hadaeghi, M. H. Zhang, T. S. Jiang, A. Zintler, L. Carstensen, Y. X. Zhang, H. J. Kleebe, H. Bin Zhang, and L. Molina-Luna, Translational Antiphase Boundaries in $NaNbO_3$ Antiferroelectrics, ACS Appl Mater Interfaces **15**, 59964 (2023).

[13] G. van Tendeloo, S. Amelinckx, C. Manolikas, and W. Shulin, The Direct Observation of "Discommensurations" in Barium Sodium Niobate (BSN) and Its Homologues, Physica Status Solidi (a) **91**, 483 (1985).

[14] H. T. Stokes, D. M. Hatch, and B. , J. Campbell, *ISODISTORT, ISOTROPY Software Suite, Iso.Byu.Edu*.

[15] B. J. Campbell, H. T. Stokes, D. E. Tanner, and D. M. Hatch, ISODISPLACE: A web-based tool for exploring structural distortions, J Appl Crystallogr **39**, 607 (2006).

[16] G. Kresse and J. Furthmüller, Efficient iterative schemes for ab initio total-energy calculations using a plane-wave basis set, Phys Rev B **54**, 11169 (1996).

[17] R. Jinnouchi, F. Karsai, and G. Kresse, On-the-fly machine learning force field generation: Application to melting points, Phys Rev B **100**, (2019).

[18] S. C. Chae, N. Lee, Y. Horibe, M. Tanimura, S. Mori, B. Gao, S. Carr, and S. W. Cheong, Direct observation of the proliferation of ferroelectric loop domains and vortex-antivortex pairs, Phys Rev Lett **108**, (2012).

[19] F. T. Huang, Y. Li, F. Xue, J. W. Kim, L. Zhang, M. W. Chu, L. Q. Chen, and S. W. Cheong, Evolution of topological defects at two sequential phase transitions of $Nd_2SrFe_2O_7$, Phys Rev Res **3**, (2021).

[20] P. Hirel, P. Marton, M. Mrovec, and C. Elsässer, Theoretical investigation of {110} generalized stacking faults and their relation to dislocation behavior in perovskite oxides, Acta Mater **58**, 6072 (2010).

[21] M. J. Watts, S. R. Yeandel, R. Smith, J. Michael Walls, and P. M. Panchmatia, *Atomistic Insights of Multiple Stacking Faults in CdTe Thin-Film Photovoltaics: A DFT Study*, in *2018 IEEE 7th World Conference on Photovoltaic Energy Conversion (WCPEC) (A Joint Conference of 45th IEEE PVSC, 28th PVSEC & 34th EU PVSEC)* (IEEE, 2018), pp. 3884–3887.

[22] Y. Wang, T. Wang, H. Arandiyan, G. Song, H. Sun, Y. Sabri, C. Zhao, Z. Shao, and S. Kawi, Advancing Catalysts by Stacking Fault Defects for Enhanced Hydrogen Production: A Review, Advanced Materials **36**, (2024).

[23] T. Z. Khan, T. Kirk, G. Vazquez, P. Singh, A. V. Smirnov, D. D. Johnson, K. Youssef, and R. Arróyave, Towards stacking fault energy engineering in FCC high entropy alloys, Acta Mater **224**, (2022).

[24] J. Li et al., Nanoscale stacking fault–assisted room temperature plasticity in flash-sintered $TiO_2$, Sci Adv **5**, (2019).

[25] R. W. Hertzberg, R. P. Vinci, and J. L. Hertzberg, *Deformation and Fracture Mechanics of Engineering Materials* (Wiley, 2012).

[26] G. van Tendeloo, J. van Landuyt, P. Delavignette, and S. Amelinckx, Compositional changes associated with periodic antiphase boundaries in the initial stages of ordering in Ni3Mo1. I. Crystallographic Analysis, Physica Status Solidi (a) **25**, 697 (1974).

[27] G. van Tendeloo, P. Delavignette, R. Gevers, and S. Amelinckx, A study of dissociated antiphase boundaries in Ni3Mo, Physica Status Solidi (a) **18**, 85 (1973).

[28] M. Lilienblum, T. Lottermoser, S. Manz, S. M. Selbach, A. Cano, and M. Fiebig, Ferroelectricity in the multiferroic hexagonal manganites, Nat Phys **11**, 1070 (2015).

[29] M. E. Holtz, K. Shapovalov, J. A. Mundy, C. S. Chang, Z. Yan, E. Bourret, D. A. Muller, D. Meier, and A. Cano, Topological Defects in Hexagonal Manganites: Inner Structure and Emergent Electrostatics, Nano Lett **17**, 5883 (2017).

[30] S. J. McCartan, P. W. Turner, J. A. McNulty, J. R. Maguire, C. J. McCluskey, F. D. Morrison, J. M. Gregg, and I. Maclaren, Anisotropic, meandering domain microstructure in the improper ferroelectric $CsNbW_2O_9$, APL Mater **8**, (2020).

[31] G. F. Nataf, M. Guennou, J. M. Gregg, D. Meier, J. Hlinka, E. K. H. Salje, and J. Kreisel, Domain-wall engineering and topological defects in ferroelectric and ferroelastic materials, Nature Reviews Physics **2**, 634 (2020).

[32] J. Seidel, Nanoelectronics based on topological structures, Nat Mater **18**, 188 (2019).

[33] G. Ren, P. Omprakash, X. Li, Y. Yun, A. S. Thind, X. Xu, and R. Mishra, Polarization pinning at antiphase boundaries in multiferroic $YbFeO_3$, Chinese Physics B **33**, (2024).

# Supplementary Information for: The Microscopic Structure of Stacking Faults in Sr$_2$NaNb$_5$O$_{15}$

Robin Sjökvist[1], Yining Xie[1], Zabeada Aslam[2], Andy P. Brown[2,3], Nicholas C. Bristowe[4], Mark S. Senn[5], Richard Beanland[1]

[1]Department of Physics, University of Warwick, Coventry CV4 7AL, United Kingdom
[2]School of Chemical and Process Engineering, University of Leeds, Leeds LS2 9JT, United Kingdom
[3]Bragg Centre for Materials Research, University of Leeds, Leeds LS2 9JT, United Kingdom
[4]Centre for Materials Physics, Durham University, Durham DH1 3LE, United Kingdom
[5]Department of Chemistry, University of Warwick, Coventry, CV4 7AL, United Kingdom

## Processing of HRTEM micrographs

It is often necessary to process HRTEM micrographs in order to remove noise and enhance contrast. The process used in this paper is outlined in Figure SI 1, where (a) shows a section of a micrograph as recorded at the microscope. In Figure SI 1 (b) the Fourier transform of the micrograph in (a) is shown. In order to remove the effect of low frequency noise, such as contamination on the sample surface, a high-pass filter is needed. This is achieved by masking the central spot in the Fourier transform. Then, in order to increase the contrast of the periodic structures in the image, a Bragg filter is applied by means of a grid-like mask. The final combined mask is shown in Figure SI 1 (c), and the inverse Fourier transform of this is shown in (d). Notice that the features are more distinguishable in (d) compared to (a), while the information is the same overall. Finally, a look up table (LUT) can be applied to convert the grayscale values to colour as shown in (e).

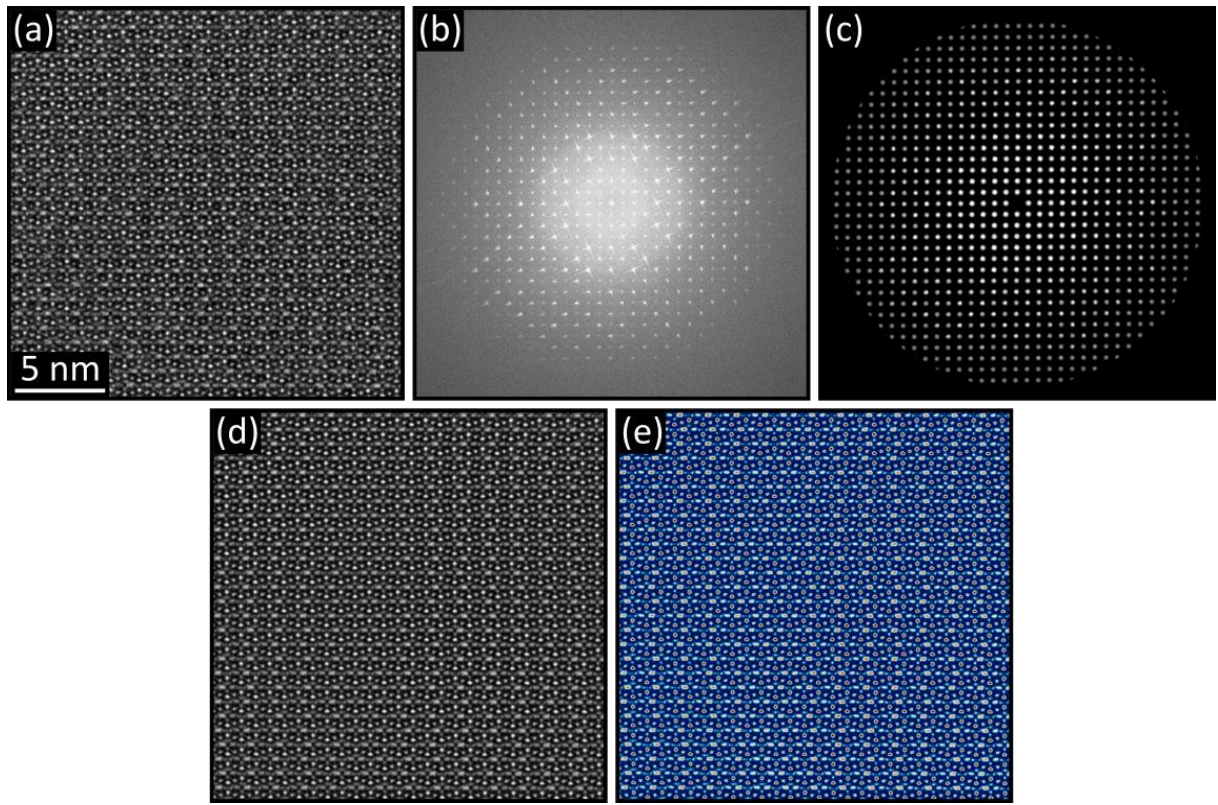

*Figure SI 1: Bragg filtering of HRTEM micrographs. (a) The as recorded micrograph. FFT, as generated (b) and after applying a mask (c). The image after filtering (d) and after applying a LUT (e).*

## Atomic resolution STEM

Atomic resolution annular dark field (ADF) STEM was carried out in the $[100]_o$ direction of SNN, in order to image the stacking fault with an alternative method. In STEM, the contrast is more related to the atomic number of the species, and the contrast observed is more directly comparable to true atomic positions than in HRTEM. The results can be seen in Figure SI 2, where a colour scheme has been applied to increase the visibility of the features, and the atomic structure of SNN is overlaid on the ADF STEM micrograph. Note that the "blue" columns show an alternating behaviour, where every other column has all sites filled and every other has half the sites filled. This gives the placement of the corners of the unit cell drawn in the centre of the figure. However, at the fault, two "half-filled" columns sit next to each other, shifted by the fault vector.

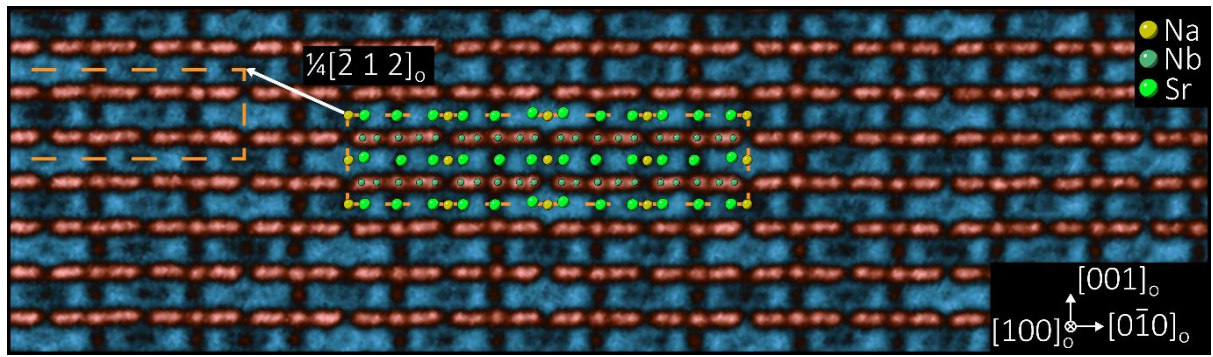

Figure SI 2: ADF STEM along the $[100]_o$ zone axis, showing atomic contrast. A colour scheme has been applied to the micrograph for visibility, and the atomic structure of SNN has been overlain to illustrate the atomic positions more clearly. The fault vector is illustrated by the white arrow, and (part of) one unit cell on either side of the fault has been indicated by an orange dashed outline.

# Atomic displacements

Full quantified atomic displacement plots showing the magnitude of movement of atoms between the constructed stacking fault structure and the annealed structure are illustrated in Figure SI 3 on a unified scale. Movement of less than 0.3 Å is seen in the majority of cases, as illustrated in Figure 5 of the main text, with the exception being a few Sr atoms at SFs that exhibit movement of around 0.6 Å.

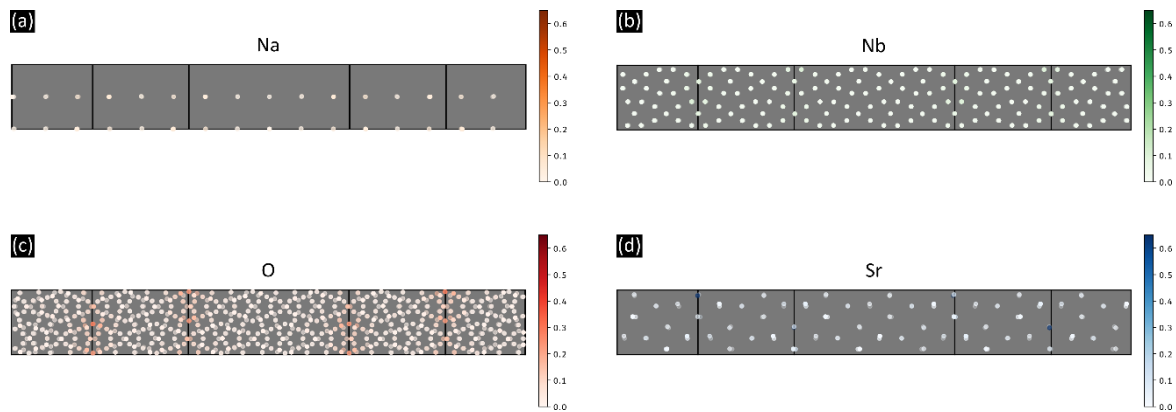

Figure SI 3: Displacements of atoms between the stacking fault structure and the resulting relaxed structure from the force-field calculations. Results for Na (a), Nb (b), O (c) and Sr (d).

# Fault frequency

Dark field TEM micrographs were studied in order to find the frequency of stacking faults in SNN. The investigation was carried out in the $[100]_o$ direction, since this is the viewing direction where the contrast from the faults is the most prominent. Figure SI 4 shows one region investigated for statistics, where the overlay indicates the faults identified. The study found that the average distance between faults in this region was around 29 nm. Since there appears to be larger gaps between individual groups of four, the median might give a better indication of the distance between faults in one such group. The result showed that the median is around 26 nm, a very similar result. This equates to a fault roughly every 7-8 unit cells along the b-axis of the orthorhombic cell.

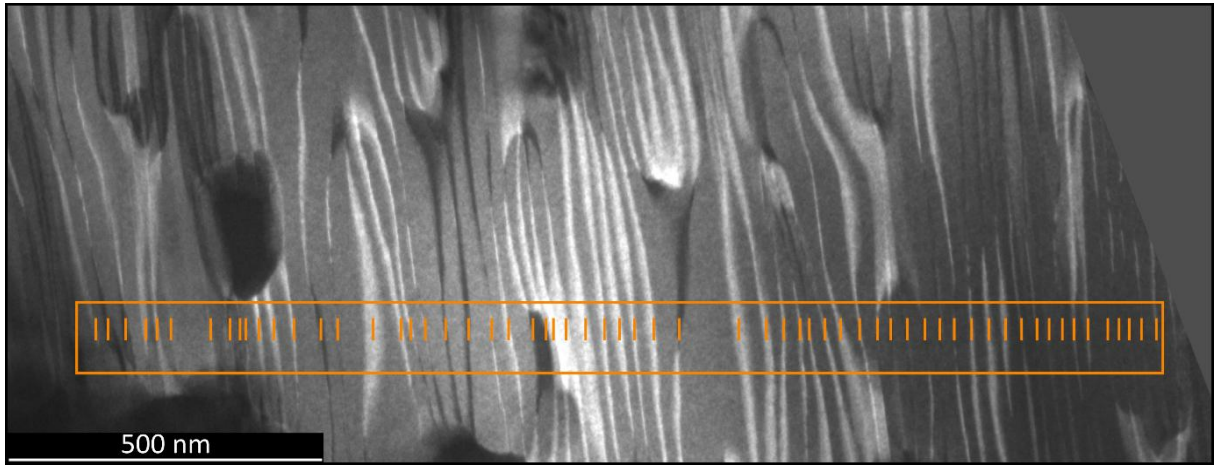

*Figure SI 4: Fault frequency in SNN. The orange box illustrates the region investigated, where orange dashes are drawn for each fault encountered.*

## Polar displacements in NbO$_6$ octahedra and symmetry mode analysis

Polar displacements of Nb atoms within NbO$_6$ octahedra were compared between the 4 unit cells wide structure in the *Cc* setting and the structure that was relaxed after the addition of stacking faults. The overall displacement is diminished along the orthorhombic c-axis with the addition of faults, as shown in the comparison in Figure SI 5. Along the a-axis, on the other hand, the displacement is seen to increase with the addition of the faults, as shown in Figure SI 6. These results are summarized in Table SI 1.

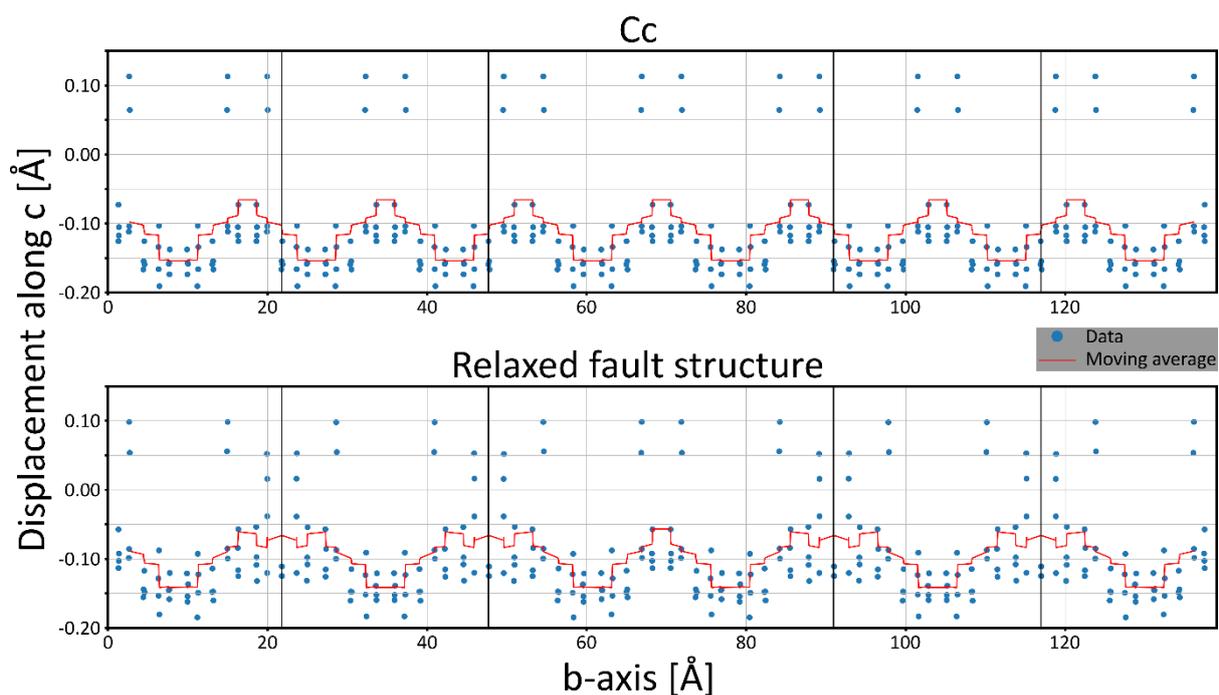

*Figure SI 5: Comparison between the Nb atomic displacements along c in the expanded cell Cc structure and the relaxed fault structure. Black lines indicate the placement of the faults.*

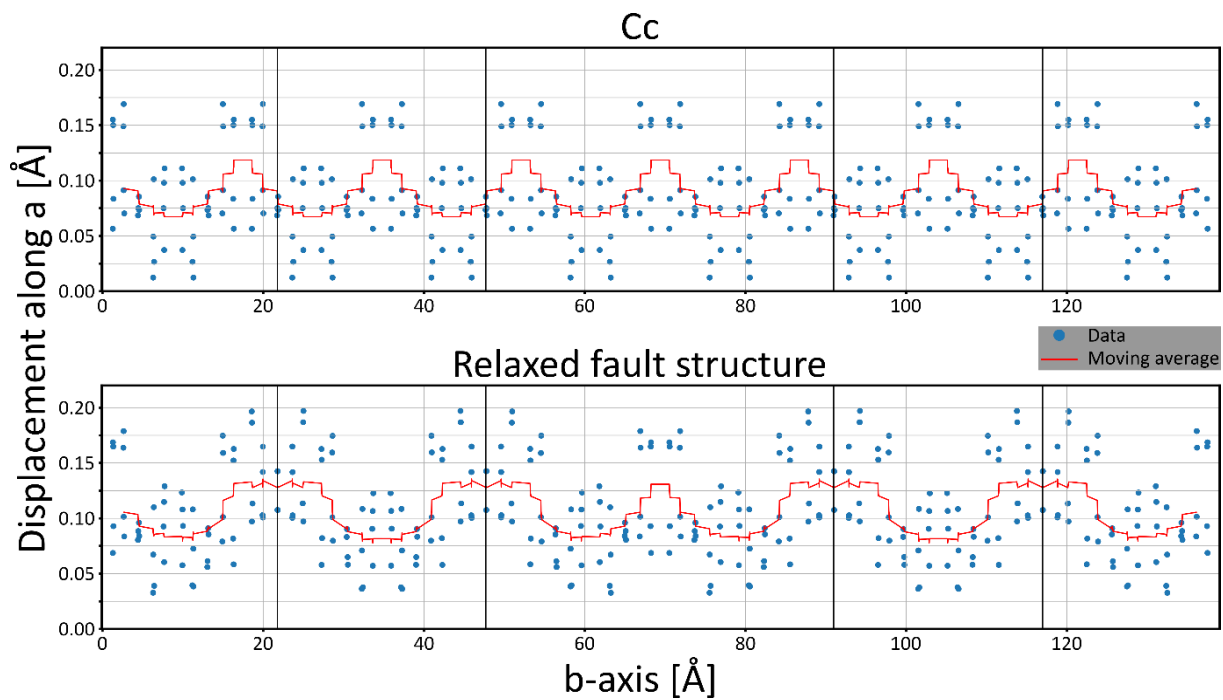

*Figure SI 6: Comparison between the Nb atomic displacements along a in the expanded cell Cc structure and the relaxed fault structure. Black lines indicate the placement of the faults.*

*Table SI 1: Summary table comparing the atomic displacements along the c and a axes of the expanded cell Cc structure and the relaxed fault structure. The percentual change due to the addition of the faults is also shown.*

| Displacement within width of fault | *Cc* structure | Relaxed fault structure | Change |
|---|---|---|---|
| **Along c** | -0.0952 | -0.0641 | -32.7% |
| **Along a** | 0.0860 | 0.1289 | +49.9% |

The relaxed structure was also cut up into 20 unit cell-sized pieces to be further analysed using ISODISTORT, as an alternative view to the atomic displacement analysis discussed previously. In this case the relevant symmetry modes to be analysed was the $\Gamma_3^-$ (polarisation along c), $\Gamma_5^-$ (polarisation along a) and $S_3$ (in-plane octahedral tilting). The results are shown in Figure SI 7, where a comparison to the average value from the analysis and the value extracted for the *Cc* structure are also shown. Table SI 2 shows a breakdown of the values.

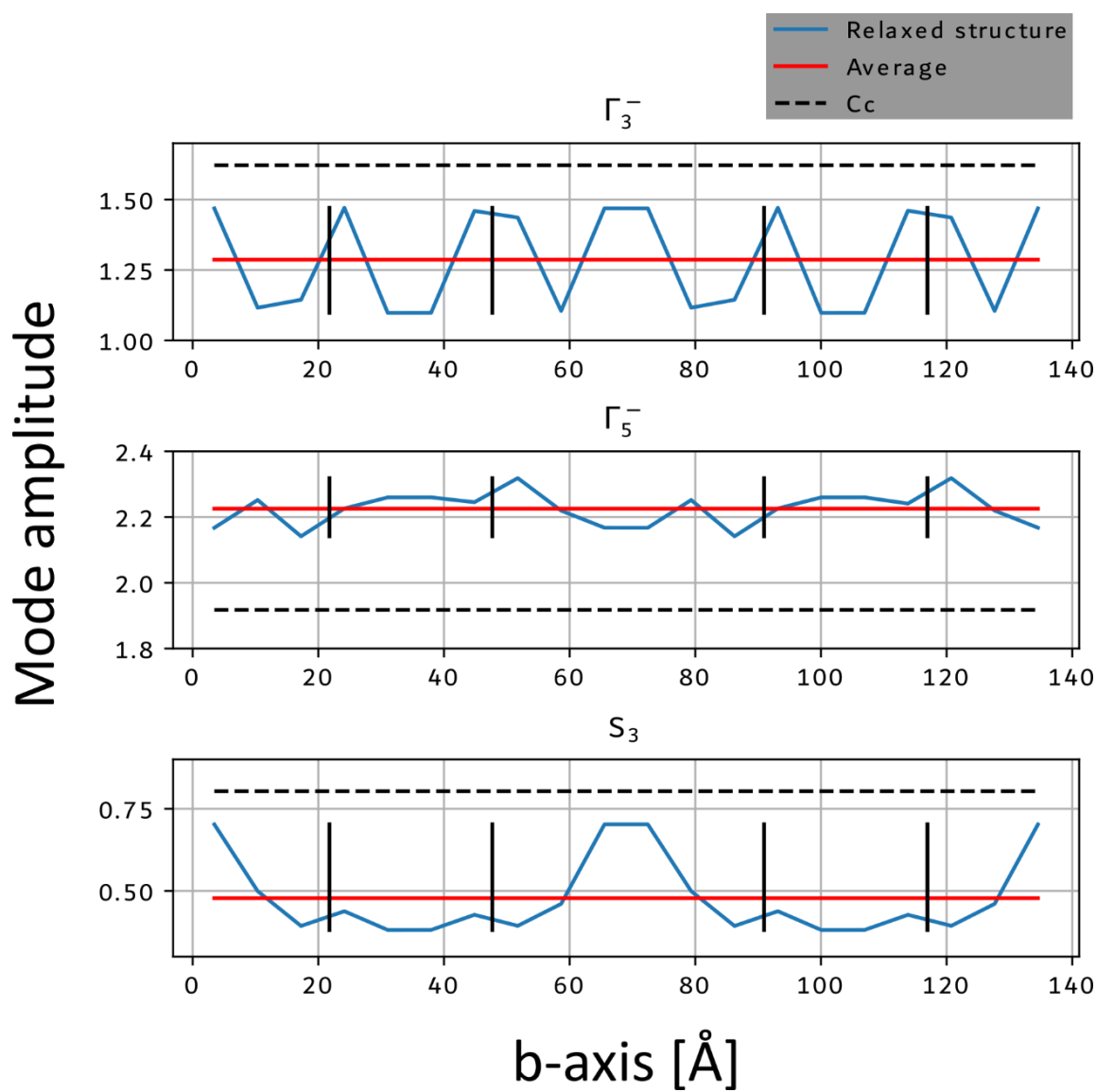

*Figure SI 7: Symmetry mode analysis of the three relevant modes $\Gamma_3^-$, $\Gamma_5^-$ and $S_3$. Black lines indicate the placement of the faults.*

Table SI 2: Comparison between the relevant symmetry mode amplitudes of the Cc structure and the relaxed fault structure. The percentual change due to the addition of the faults is also shown.

| Average mode amplitude | Cc structure | Relaxed fault structure | Change |
|---|---|---|---|
| **GM3-** | 1.622 | 1.287 | -20.7% |
| **GM5-** | 1.918 | 2.226 | +16.1% |
| **S3** | 0.804 | 0.478 | -40.5% |

# 4D-STEM

4D-STEM was carried out in order to confirm the presence of polar regions in the material that do not coincide with the domains formed by the stacking faults. Figure SI 8 highlights the results from the 4D-STEM investigation. In (a), the summed diffraction pattern from the entire region is shown. In order to construct the virtual dark field image displayed in (b), the circled spot in Figure (a) was selected in post processing. This selectively highlights the regions of a particular out-of-plane polarization in the image, making them bright, while the regions of opposite polarization will become dark. Then, a virtual aperture is placed in the dark field image to generate average diffraction patterns from limited regions corresponding to one or the other polarization, as shown in panels (c) and (d). The arrows in panel (b) point to four fold meeting points of stacking faults, illustrating that no drastic contrast change is associated with the faults.

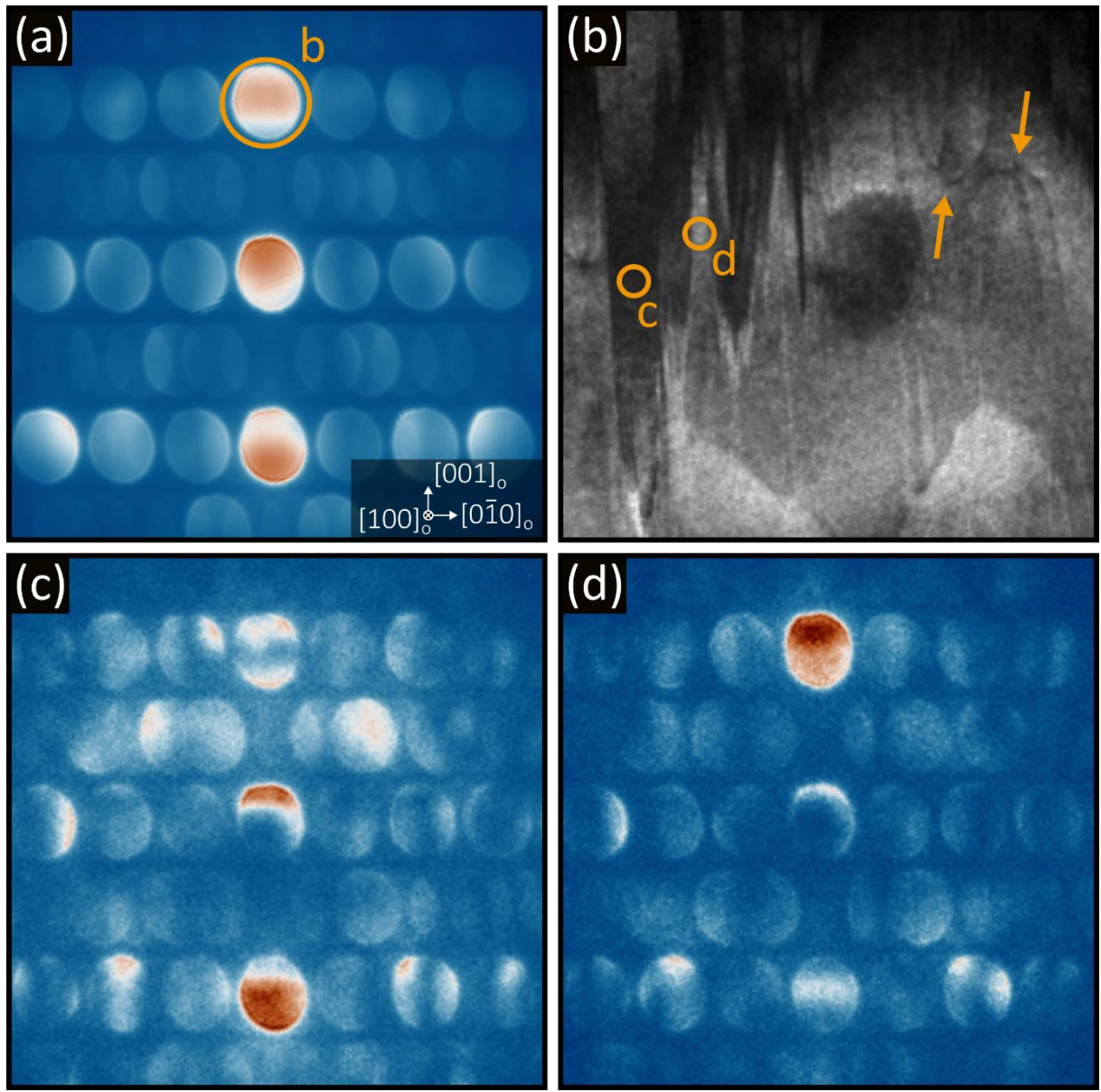

*Figure SI 8: 4D-STEM data of SNN. The full, average, diffraction pattern (a) of the investigated region, shows the spot used to construct the virtual dark field image (b). In (b), virtual apertures are used to construct the diffraction patterns from regions related to downwards (c) or upwards (d) polarization. The orange arrows in (b) point to four fold meeting points of stacking faults.*

Furthermore, Figure SI 9 illustrates a case where a polar domain stretches into a twin domain. Figure SI 9 (a) shows a 4DSTEM bright field image, illustrating polar domains of alternating intensity. Virtual dark field was constructed by selecting only the rows which contain the ¼ order superstructure spots as discussed in Figure 2 of the main text. This allows a clear separation between twins oriented along the orthorhombic [100] direction (bright) and twins oriented in the orthorhombic [010] direction (dark). The arrow points to a region where a polar domain, as evidenced in (a), stretches into the twin domain.

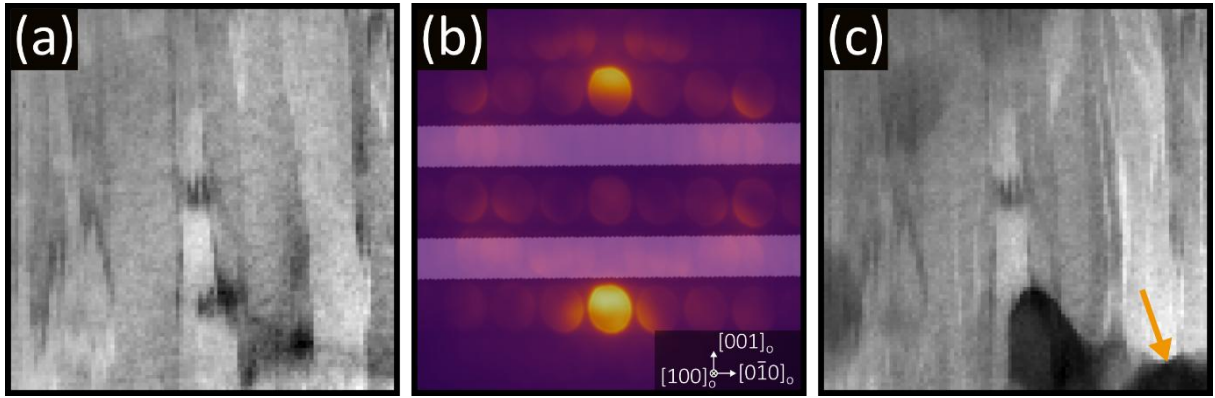

*Figure SI 9: Polar domains that are separate from the twin domains. (a) shows a virtual bright field micrograph generated from a 4D STEM dataset, where stripes of alternating contrast represent polar domains. (b) shows the mask used to construct the virtual dark field presented in (c).*